\documentclass[12pt]{iopart}

\newcommand{\bea}{\begin{eqnarray}}
\newcommand{\eea}{\end{eqnarray}}
\usepackage{graphicx}
\usepackage{amssymb}

\begin{document}
\title[Fundamental statistical features and self-similar properties of tagged networks]
{Fundamental statistical features and self-similar properties of tagged networks}
\author{Gergely Palla$^1$, Ill\'es J. Farkas$^{1}$, P\'eter Pollner$^1$, 
Imre Der\'enyi$^2$ and Tam\'as Vicsek$^{1,2}$}
\address{$^1$ Statistical and Biological Physics Research Group of HAS and}
\address{$^2$ Dept. of Biological Physics, E\"otv\"os Univ.,
  1117 Budapest, P\'azm\'any P. stny. 1A}

\begin{abstract}
We investigate the fundamental statistical features of tagged 
 (or annotated) networks having a rich variety of attributes associated
with their nodes. Tags (attributes, annotations, properties, features, etc.)
provide essential information about the entity represented by a given node,
 thus, taking them into account represents a significant step towards 
a more complete
description of the structure of large complex systems. Our main goal here is
to uncover the relations between the statistical properties of the node tags
and those of the graph topology.
In order to better characterise the networks with tagged nodes, we
introduce a number of new notions, including tag-assortativity (relating
link probability to node similarity), and new quantities, such as
node uniqueness (measuring how rarely the tags of a node occur in the
network) and tag-assortativity exponent.
We apply our approach to three large networks
representing very different domains of complex systems. A number of the
tag related quantities display analogous behaviour (e.g., the networks we
studied are tag-assortative, indicating possible universal aspects of tags
versus topology), while some other features, such as the distribution of the
node uniqueness, show variability from network to network allowing for
pin-pointing large scale specific features of real-world complex networks.
We also find that for each network the topology and the tag
distribution are scale invariant, and this self-similar property of
the networks can be well characterised by the tag-assortativity
exponent, which is specific to each system.
\end{abstract}
\maketitle

\section{Introduction}
\label{sec:intro}

Many complex systems in nature and society can be successfully represented
in terms
of \emph{networks} capturing the intricate web of connections among the units
they are made of \cite{Laci_revmod,Dorog_book}. 
In the recent
 years, the research in this field have been focused mainly on the
 \emph{topology} of the graphs corresponding
 to these real networks.
Since
 this approach is rooted in, among others, 
statistical physics, where often 
the thermodynamic limit is considered and also the size of the known nets
 is becoming huge, 
several {\it large-scale} properties of real-world 
webs have been uncovered, 
e.g., a low average distance 
combined with a high average clustering coefficient \cite{Watts-Strogatz},
the broad (scale-free) distribution of node degree 
(number of links of a node)
\cite{Faloutsos,Laci_science,Boccaletti,Jeong_metabolic}
and various signatures of hierarchical/modular organisation
\cite{Ravasz02,Han04}. 

On the other hand, there has been a 
quickly growing interest in the \emph{local structural units} 
of networks. Small and well defined sub-graphs consisting
of a few vertices have been introduced as \emph{motifs} \cite{Alon_1,Alon_2}, 
whereas somewhat larger units, associated with more highly interconnected parts
\cite{Domany_PRL,GN-pnas,Zhou,Newman_fast,Radicchi,Huberman_PNAS,Reichard-Bornhold_PRL,Scott_book,pnas-suppl,Everitt-book,Knudsen-book,Newman-europhys,
CPM_nature,Fortunato_coms_chap,Fortunato_coms_cikk}
 are usually called \emph{communities},
clusters, cohesive groups, or modules.  These structural 
 sub-units can correspond to multi-protein functional units in
molecular biology \cite{Ravasz02,Spirin_PNAS},
a set of tightly coupled stocks or industrial
sectors in economy \cite{Onnela-taxonomy,Saramaki_stock_Phisica_A},
groups of people \cite{Scott_book,Watts-Dodds,group_evolv_nature},
 cooperative players \cite{Szabo_PRE1,Szabo_PRE2,Szabo_Phys_Reports},
 etc. The
location of such building blocks can be crucial to the understanding of
the structural and functional properties of the systems under investigation.

The majority of the complex network studies concern ``bare'' graphs 
corresponding to
a simple list of connections between the nodes, or at most 
weighted networks where a connection strength (or intensity) is associated
 to the links. However, the introduction of \emph{node tags} 
(also called as attributes, annotations, properties, categories, features) 
 leads
 to a richer structure, opening up the possibility for a more comprehensive
 analysis of the systems under investigation. 
These tags can correspond to basically
 any information about the nodes and in most cases a single node 
 can have several tags at the same time. The use of such annotations
 in biological networks is a common practice \cite{Mason_nets_in_bio,Zhu_nets_in_bio,Aittokallio_nets_in_bio,Finocchiaro_cancer,Jonsson_Bioinformatics,Jonsson_BMC}, where the tags usually refer to the biological function of 
 the units (proteins, genes, etc.).
Another interesting
 application of node features can be seen in the studies of co-evolving
 network models, where the evolution of the network topology affects
 the node properties and vice versa 
\cite{Zimmermann_coevlov,Eguiluz_coevolv,Watts_science,Ehrhardt_coevolv,Newman_coevolv,Gil_coevolv,Vazquez_PRE,Vazquez_cond_mat,Kozma_coevolv,Benczik_coevolv}. These models are aimed at describing
 the dynamics of social networks, in which people with similar opinion 
 are assumed to form ties more easily, and the opinion of connected people
 becomes more similar in time. Finally, we mention the study of collaborative
 tagging in Ref.\cite{Lambiotte_tags}, where tripartite networks were 
constructed from data concerning users who associated tags
 to some kind of items, (such as music listeners  
classifying music records). 
The three types of nodes corresponded to the users, 
the tags, and the items. The tagging was
 carried out without any central authority and according to the results, 
 the analysis of the bi- and unipartite projection of the networks can help
 in structuring the contents (e.g., define a hierarchy between the tags).

In this paper we study tagged networks from yet another point of view.
Our focus is on networks where the links are in principle 
not related to tagging, however tags can be associated with the
 nodes quite naturally.
The PACS numbers or key-words
 in case of co-authorship networks, the scope of business or the industrial
 sector of companies in the context of financial networks, or the status of
 employees in the case of a network representing the social ties inside
 a large firm provide plausible examples for possible tags. 
The complexity of the networks studied these days is rapidly increasing
 together with their size. The use of tags associated with the nodes
 can help in revealing hidden structures or fasten searching within the
 networks. Since the 
usefulness of such attributes has already been proven in biology, 
the inclusion of tags in the analysis of other networks as well
 is expected to give a deeper understanding of the interrelations 
 shaping the structure and dynamics of the systems under study.

Along this line, 
in the present paper we study the fundamental statistics characterising
the distribution of tags in large
 annotated real networks.
 By choosing networks representing
 completely unrelated systems (a co-authorship network, a
 protein interaction network, and the English Wikipedia),
 we seek for signs of universality in these statistics. Furthermore,
 we are interested in the relations between the network topology and the 
distribution of tags.
The tags enable the definition of a \emph{similarity function} between
 the nodes which is a priori independent of the topology. We shall
 refer to this quantity as the \emph{tag-similarity} of the nodes in
 order to distinguish it from the usual structural similarity of the
 nodes (based on the similarity between the nearest neighbours).
 The study of the tag-similarity opens up
 further directions for exploring the intricate relations between the
 annotations and the graph structure itself. Interestingly, in all selected 
systems, the tags form a sort of \emph{taxonomy}: they correspond to
features ranging from very specific to rather general ones, which are
embedded in a hierarchic structure held together by 
``is a sub-category of'' type relations. This inter-relatedness of the 
 tags adds an extra twist to the definition of the quantities we study. 

The paper is organised
 as follows. In Sect.\ref{Sect:defi}. we define the most important
 quantities we aim to study, whereas the construction of the investigated
 networks (and the hierarchy of the corresponding node labels) is 
 detailed in Sect.\ref{Sect:construct}. The results are presented in 
Sect.\ref{Sect:results} and we close the paper with with some
 concluding remarks in Sect.\ref{Sect:summary}.

\section{Definitions}
\label{Sect:defi}

\subsection{Basic statistics}

\subsubsection{Number of tags on a node}

In principle, nodes in a network can be tagged with almost anything.
 Here we list a few basic types followed by particular examples in
 parenthesis: real numbers (the accumulated impact factors of authors in a 
co-authorship network), integers (the number of articles of an author),
 or character strings (functions of proteins in a protein-protein interaction network). However,
 in most cases, (including the systems we study in the present paper), 
the node attributes correspond to \emph{character strings},
 chosen from a finite set of possible tags. Usually a node can
 have more than one tag attached to it, e.g., numerous proteins appearing
 in a protein-protein interaction (PPI) network  have 
 multiple functions. One of the basic statistics
 about the annotations is the distribution of the number of tags
 on the nodes. 

\subsubsection{Tag frequencies}

\begin{figure}[t!]
\centerline{\includegraphics[angle=0,width=\columnwidth]{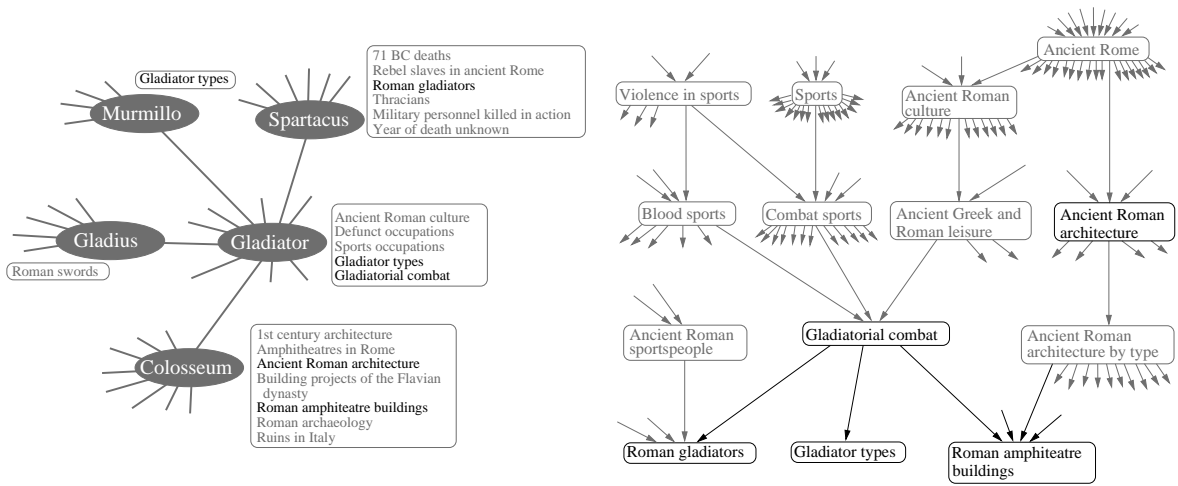}}
\caption{A small labelled sub-graph and the corresponding DAG of categories in
 the English Wikipedia.
In the left panel  we show a few neighbours of the page ``Gladiator'', 
where the 
connections correspond to mutual hyper links between the pages embedded in
 the text of the page. At the bottom of each page we can find a list 
 of categories, which we use as tags. These are listed in the frames
 appearing near the nodes. These categories are organised into a 
 DAG, as demonstrated in the right panel, 
where e.g., ``Gladiator types'' 
is a sub-category  of ``Gladiatorial combat''. The categories appearing in 
both 
panels are emphasised in black.}
\label{fig:Wiki_DAG_illustrate}
\end{figure}

Similarly to the varying number
 of tags on the nodes, the \emph{frequency} of the different tags can also be
 rather heterogeneous. What makes the picture even 
 more complex is that in many cases the tags refer to
 \emph{categories} 
of a \emph{taxonomy} or \emph{ontology} (capturing the view of a certain domain, e.g., protein functions). This means that the tags are 
organised into a structure of relationships
which can be represented by a directed acyclic graph (DAG), where the 
 directed links between two categories represent an ``is a sub-category of'' 
relation. The nodes close to the root in the DAG are usually related to
 general properties, and as we follow the links towards the leafs, the
 categories become more and more specific.
 In some cases we can find categories in the DAG with more than one 
in-neighbours, meaning that the given sub-category is part of more than
 one categories (that are not parts of one another). 
Also note that nodes can be classified not only by
the leaf-categories e.g., several proteins in a PPI network can be
 found with rather general functional descriptions.
We illustrate the concept of tagged networks and the corresponding DAG of
categories in Fig.\ref{fig:Wiki_DAG_illustrate}. with the help of 
 the English Wikipedia.

Given the DAG between the possible tags, we can define 
the frequency of a given tag $\alpha$ in two 
different ways:
\numparts
\begin{eqnarray}
p_{\alpha}&\equiv& N_{\alpha}/N, \\
\tilde{p}_{\alpha}&\equiv& \tilde{N}_{\alpha}/N,
\label{eq:tag_freq}
\end{eqnarray}
\endnumparts
where $N_{\alpha}$ denotes the number of nodes tagged with $\alpha$, 
$\tilde{N}_{\alpha}$ stands for the number of nodes tagged with $\alpha$ or
any of its descendents, and $N$ is equal to the total number of nodes in 
 the network. From these definitions it follows that when the
 number of un-tagged nodes is zero, the root of the annotation DAG will receive
 $\tilde{p}_{\alpha}=1$, whereas for the leaf categories 
$\tilde{p}_{\alpha}=p_{\alpha}$. Furthermore, if category $\beta$ is a 
descendent of $\alpha$, then $\tilde{p}_{\alpha}\geq\tilde{p}_{\beta}$. 
Low frequency tags are more specific in an information theoretical sense,
whereas high frequency tags carry almost no information (e.g., being
 tagged by the root in the annotation DAG adds absolutely no information
 to the description of a node).

In the following, we shall refer to the sub-graph induced by the nodes
 (i.e. constituted by these nodes and all links between them) marked
 by the tag $\alpha$ and any of its descendents as the 
\emph{tag-induced sub-graph} of $\alpha$. The number of nodes in this sub-graph
 is given by $\tilde{N}_{\alpha}$, whereas the number of links can vary between
 $\tilde{M}_{\alpha}=0$ and 
$\tilde{M}_{\alpha}=\tilde{N}_{\alpha}(\tilde{N}_{\alpha}-1)/2$. It is interesting
 to compare $\tilde{M}_{\alpha}$ to the number of links $\tilde{M}_{\rm rand}$
one would expect in a random  sub-graph of the same size: 
if $\tilde{M}_{\alpha}$ is significantly larger/smaller 
than $\tilde{M}_{\rm rand}$, then nodes sharing the tag $\alpha$ attract/repel
 each other in the sense that they are
 linked with higher/smaller probability than at random.

\subsection{Tag-similarity}
Our aim in this section is to define a similarity function between the nodes
 which is based solely on the tags, therefore, it can be evaluated without
 any knowledge about the graph structure. Although we refer to this quantity
 as the tag-similarity of the nodes in general, we shall
 use the term similarity in the same sense for short.

\subsubsection{Simple similarity measures}
To what extent two nodes $i$ and $j$ having a set of tags $\Omega_i$ and
 $\Omega_j$ are similar  
is a question far from trivial, as the number of possible similarity 
measures is vast. A simple approach is to use the Jaccard-similarity 
\cite{Jaccard} defined as 
\begin{equation}
s_{ij}^{\rm (J)}\equiv\frac{\left|\Omega_i\cap\Omega_j\right|}{\left|\Omega_i\cup\Omega_j
\right|},
\label{eq:Jaccard}
\end{equation}
where $\left|\Omega_i\cap\Omega_j\right|$ is equal to the number of
 common tags and $\left|\Omega_i\cup\Omega_j\right|$ is equal to the total
 number of different tags in $\Omega_i$ and $\Omega_j$. Another possibility
 is to represent the annotations as vectors $\mathbf{v}_i$ and 
$\mathbf{v}_j$, where the 
number of entries
 in the vectors is equal to the number of different tags in the network, 
and the non-zero elements indicate the presence (or possibly the weight) of 
 the actual tags on the given node.
 In this approach the cosine similarity 
\begin{equation}
s_{ij}^{\rm (c)}\equiv\frac{\mathbf{v}_i\cdot\mathbf{v}_j}{\left|\mathbf{v}_i\right|
\left|\mathbf{v}_j\right|}
\label{eq:cosine}
\end{equation}
yields a simple similarity value for a pair of nodes $i$ and $j$.

The advantage of the above methods is that they do not depend on the DAG
 between the tags, therefore, they can be applied even when the
 tags are not part of a structured taxonomy. However, when a tag refers
 to a sub-category of another tag, the similarity measure should be refined.
 As an example, let us assume that node $i$ is tagged  exclusively with 
 category $\alpha$, and node $j$ has a single tag $\beta$, that
 is a direct descendent of $\alpha$ (e.g., $\alpha\equiv$''knife'' 
 and $\beta\equiv$''kitchen knife''). In this case both (\ref{eq:Jaccard})
 and (\ref{eq:cosine}) yield $s^{\rm (J)}_{ij}=s_{ij}^{\rm (c)}=0$, 
which is not what we would expect. 

\subsubsection{Semantic similarities}
To overcome the problem raised above, we should use a similarity measure
 which takes into account the structure of the annotation DAG. At this 
 point we divide the evaluation of similarity into two parts:
 first we deal with the similarity $s_{\alpha\beta}$ between a 
 pair of tags, then elaborate on how to combine the pairwise similarities
 $s_{\alpha\beta},\alpha\in\Omega_i,\beta\in\Omega_j$ between the sets of 
tags $\Omega_i$, $\Omega_j$ associated with a pair of 
nodes $i$ and $j$ to obtain $s_{ij}$. 

A simple choice for determining
 the similarity between two tags is the length of the longest
 shared path towards the root of the annotation DAG. 
A somewhat more sophisticated
 approach is to use \emph{semantic similarities}. The basic idea behind these
 methods is to take into account the frequency of the tags:
sharing a rare tag by two nodes should indicate high similarity, 
whereas sharing a frequent tag should not. The 
semantic similarity between tags $\alpha$ and $\beta$ derived by
 Resnik \cite{Resnik} as
\begin{equation}
s_{\alpha\beta}^{\rm (R)}\equiv\max_{\gamma\in \Gamma(\alpha,\beta)}\left[ -\log \tilde{p}_{\gamma}\right],
\label{eq:Resnik}
\end{equation}
where $\Gamma(\alpha,\beta)$ denotes the set of common  ancestors of
 $\alpha,\beta$, and $-\log \tilde{p}_{\gamma}$ corresponds to the 
 information content of category $\gamma$. From this definition it follows
 that if $\beta$ is a descendent of $\alpha$, then 
$s_{\alpha\beta}^{\rm (R)}=-\log \tilde{p}_{\alpha}$, and when the two compared 
 tags are not connected by a directed path, then $s_{\alpha\beta}^{\rm (R)}$
 is equal to the information content of one of their nearest common ancestors. 
A closely related similarity measure was proposed
 by Lin \cite{Lin} as 
\begin{equation}
s_{\alpha\beta}^{\rm (L)}\equiv\frac{2\max_{\gamma\in \Gamma(\alpha\beta)}\left[ 
-\log \tilde{p}_{\gamma}\right]}{\left|\log \tilde{p}_{\alpha}+\log \tilde{p}_{\beta}\right|}.
\label{eq:Lin}
\end{equation}
In practice (\ref{eq:Lin}) was reported to slightly under perform 
(\ref{eq:Resnik}) \cite{Guo}, 
however the big advantage of (\ref{eq:Lin}) is that
 $s_{\alpha\beta}^{\rm (L)}$ becomes bounded in $[0,1]$. 
The maximal possible $s_{\alpha\beta}^{\rm (R)}$ obtained 
from (\ref{eq:Resnik}) depends on the frequency of the 
rarest tag, which in our case is strongly varying from system 
 to system. For this reason, 
we shall use (\ref{eq:Lin}) for calculating the similarity between
categories.

When moving from the similarity of tags to the similarity of nodes, 
again we have a number of
 possibilities to choose from. A simple approach is to use the average
 of the pairwise similarities as
\begin{equation}
s_{ij}\equiv \frac{1}{n_in_j}
\sum_{\alpha\in \Omega_i,\beta\in\Omega_j}s_{\alpha\beta}^{\rm (L)},
\label{eq:node_sim_av}
\end{equation}
 where $n_i$ and $n_j$ denote the 
number of tags on node $i$ and $j$ respectively. The problem with
 the expression above is that if the labels associated with a given node 
are very different from each other, then by comparing this node
 even to itself, the ``cross-terms'' reduce the similarity value. 
A simple solution is to replace the average in (\ref{eq:node_sim_av})
 by the maximal pairwise similarity amongst the tags:
\begin{equation}
s_{ij}\equiv \max_{\alpha\in \Omega_i,\beta\in\Omega_j}s_{\alpha\beta}^{\rm (L)}.
\label{eq:max_node_sim}
\end{equation}
Another possibility along this line
is to organise the pairwise similarities between the tags 
into an $n_i$ by $n_j$ matrix, and define the
quantities $rowScore$ and $columnScore$ as the average of the maximal values
 in the rows and columns of this matrix respectively. The similarity
 between the two annotation vectors can then be given as either the
 average or the maximum of $rowScore$ and $columnScore$ \cite{Schlicker}.
In our studies we shall use (\ref{eq:max_node_sim}) due to its 
 computational simplicity and the fact that it is analogous 
to the concept of minimum linkage clustering, where
the distance between two sets of elements (the tags) is defined as the 
minimum pairwise distance between the elements. 

\subsection{Tag-assortativity}
A plausible hypothesis about tagged real networks is that links 
are likely to form ties between similar nodes and vice versa, 
we expect connected nodes to share common tags with enhanced probability.
However, this property is not evident in all cases. E.g., if we colour
the nodes in a network according to the famous vertex colouring problem
\cite{graph_colouring}, (namely we seek for the minimal number of
 colours which can be distributed in such a way that no neighbours have
 the same colour), and identify the node colours as the tags, then similar
 nodes are actually never connected. 

In general, the property that nodes are more frequently connected to
others that are similar/different in some quality is referred to as
assortativity/disassortativity. The most typically considered quality --
which is based on the network's topology -- is the degree of the nodes.
In tagged networks, however, another natural way of comparing nodes can
be based on the above defined tag-similarity. We can thus introduce the
notion of \emph{tag-assortativity} (to distinguish this property from
the degree-assortativity), and call a network
\emph{tag-assortative}/\emph{tag-disassortative} if nodes having
similar tags are linked with higher/lower probability than at random.

\subsection{Uniqueness}

Interestingly it is not uncommon to find tags associated with the same node
 which
 are rather different from each other, e.g., in the PPI network studied
 in this paper more than 10\% of the nodes have at least 
one pair of tags for which the nearest common ancestor is actually the root 
 of the annotation DAG. This means that the given protein can take part in
 very different biological processes.
On the other hand, many nodes have more or less similar
categories in their annotation, so they take part in more or less similar
 processes. 

To quantify the above aspect,
 we introduce the \emph{node uniqueness}, defined as
\begin{equation}
u_i\equiv \min_{\alpha,\beta\in\Omega_i}s_{\alpha\beta}^{\rm (R)}
\label{eq:K}
\end{equation}
In principle, we could have chosen $s_{\alpha\beta}^{\rm (L)}$ rather than $s_{\alpha\beta}^{\rm (R)}$ in the definition above. However,
 since $s_{\alpha,\alpha}^{\rm (L)}=1$ for every $\alpha$, if node $i$ has 
only a single tag, then $u_i$ would be unity independent of whether this
 tag is frequent or not. By using the the Resnik-similarity (\ref{eq:Resnik})
 for which $s_{\alpha,\alpha}^{\rm (R)}=-\log\tilde{p}_{\alpha}$, we can 
differentiate between nodes with single tags as well, based on the 
tag frequencies. 
The lowest possible value for $u$ occurs in the case where a node
 belongs to more than one categories, out of which at least two have the 
 root of the DAG as their nearest common ancestor. 
The highest possible value for $u$ occurs if a node
 belongs to a single category, and this category happens to be the rarest
 among all. We note that in Ref.\cite{Lambiotte_tags} a closely
 related quantity called node diversity was defined for the case where
 the tags are not part of a hierarchical taxonomy.

\section{Applications}
\label{Sect:construct}

We studied the node annotations in three networks of high importance
 from the aspect of practical applications, capturing the relations
 between interacting proteins, collaborating scientists, and pages of
 an on-line encyclopedia. The PPI network of MIPS \cite{MIPS} contained $N=4546$ proteins, 
connected by $M=12319$ links, and the tags attached to the nodes corresponded 
 to $2067$ categories describing the biological processes the proteins
 take part in. The DAG between these categories was obtained from
 the Genome Ontology database \cite{GO}. 

The investigated co-authorship network is known as the MathSciNet 
(Mathematical review collection of the American Mathematical Society)
 \cite{Mathscinet}, 
and represents
 the $M=873775$ links of collaboration between $N=391529$ mathematicians. 
The node tags were obtained from the $6499$
 different subject classes of the articles, which were organised
 into a DAG. Thus, the set of tags attached to each author was 
the union of all subject-classes that appeared on her/his papers. 

Finally, the nodes in the third studied network corresponded to the 
$N=1473894$ pages of the English Wikipedia \cite{Eng_wiki,Zlatic_wikipedia,Capocci_wiki_PRE,Capocci_wikipedia}, connected by the 
$M=3755485$ hyperlinks 
embedded in the text of the pages. At the bottom of each page, one can find
 a list of categories, which were used as node tags. Since each 
 wiki-category is a page in the Wikipedia as well, we removed these 
pages from the network to keep a clear distinction between nodes 
and attributes. Furthermore, we kept only the mutual links between 
the remaining pages. Similarly
 to the biological processes in the MIPS network or the subject
 classes in the MathSciNet, the wiki-categories can have sub-categories
 and are usually part of a larger wiki-category. However, when representing
 these relations as a directed graph, some directed loops appear, 
 therefore, they do not form a strict DAG as required for e.g., the semantic
 similarity measures (\ref{eq:Resnik}-\ref{eq:Lin}). 
In order to be able to use these similarity functions, we removed a 
few relations from this graph until it turned into a DAG,
 following a method detailed in the Appendix. 

Due to the very large size
 of this network, some of the analysis we carried out turned out to 
be very time consuming, therefore, in certain cases 
we used only smaller sub-graphs of the Wikipedia, 
induced by rather general categories e.g., ``Soccer'', ``Japan'', etc. 
(The tags which were not descendents of the chosen category were naturally 
dropped from the nodes in the tag-induced sub-graph).  
 The advantage of this method is that the categories
 appearing as node tags in the resulting sub-graph also form a DAG which 
 is equivalent to the 
 DAG of the descendents of $\alpha$ (in which the root is $\alpha$). 
In this paper we show
 the results for the case where $\alpha\equiv$``Japan'', (altogether
  $N=43307$ nodes, $M=102753$ links and $3197$ sub-categories), however other
 choices resulted in very similar results as well.

We also checked whether this sort of sampling from the networks 
distorts the studied statistics by examining tag-induced sub-graphs in
the other two networks (and smaller tag-induced sub-graphs in 
the Wikipedia/Japan network) as well. We found that for all statistics studied
 in this paper the results in a large enough tag-induced sub-graph are 
 very similar to those for the whole network, and the differences can  
 be mostly attributed to the different system sizes.

\section{Results}
\label{Sect:results}

\subsection{Basic statistics}
\begin{figure}[t!]
\centerline{\includegraphics[angle=0,width=\columnwidth]{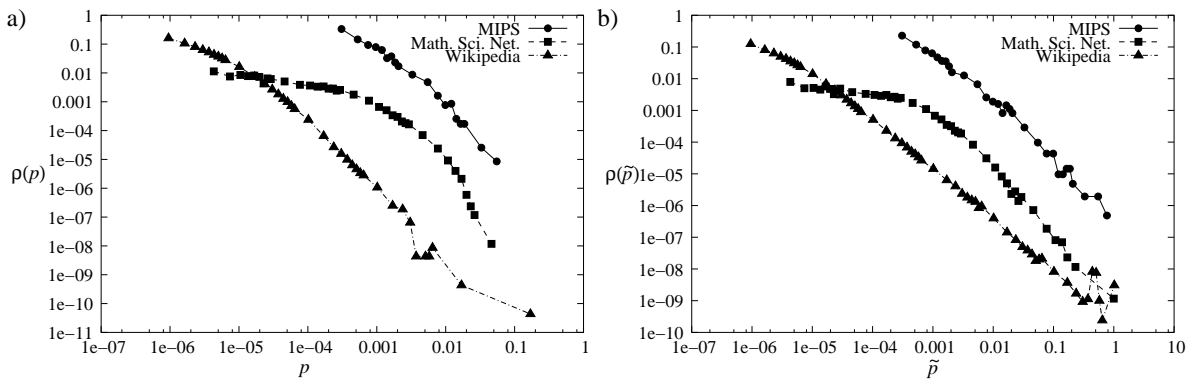}}
\caption{The density distributions of the tag frequencies a) $p_{\alpha}$ 
and b) $\tilde{p}_{\alpha}$ on logarithmic scale. }
\label{fig:cat_ocs}
\end{figure}

We begin our investigations in Fig.\ref{fig:cat_ocs}. with the distribution 
of the tag
frequencies in the three networks. According to 
Fig.\ref{fig:cat_ocs}a, the distribution
 of $p_{\alpha}$ 
resembles a power-law for the MIPS network and the Wikipedia, 
whereas it resembles an exponential distribution for the MathSciNet. 
When moving from $p_{\alpha}$ to $\tilde{p}_{\alpha}$ (by including the nodes
 tagged with any descendents of $\alpha$ as well), the tail of the distribution
 becomes power-law like for each network, as shown in Fig.\ref{fig:cat_ocs}b.
This is consistent with the hierarchical nature of the annotation DAG:
categories high up in the DAG correspond to general concepts, therefore
 apply to a vast number of nodes, whereas  leaf categories (without
 any descendents) refer to something specific, therefore occur 
rarely \cite{Rios_branching,Caldarelli_branching,Caldarelli_taxonomy}.

\begin{figure}[t!]
\centerline{\includegraphics[angle=0,width=\columnwidth]{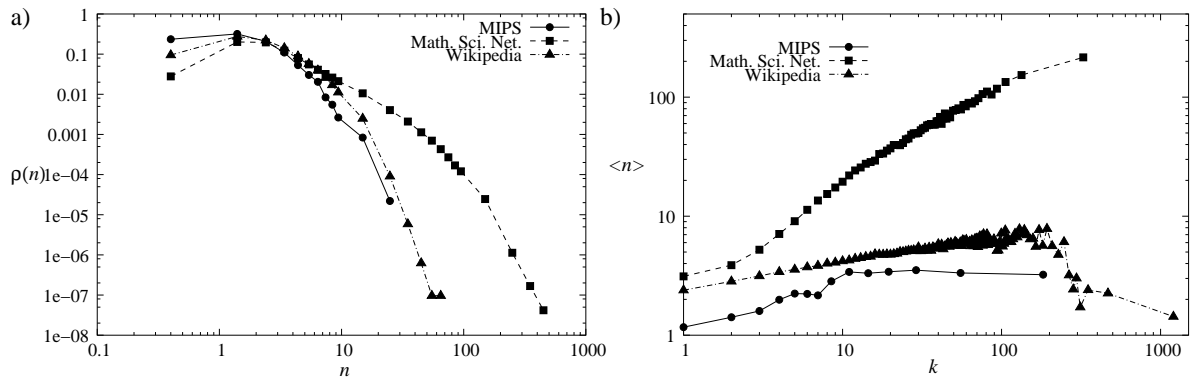}}
\caption{a) The density distributions of the number of tags $n$ per node. b)
The average number of tags $\left< n\right>$ as a function of 
the node degree $k$. }
\label{fig:num_categs}
\end{figure}
Our main goal in this paper is to study the relations between
 the distribution of node tags and the network topology. One
 of the most basic statistical quantity which can be studied in this respect
 is the number of $n_i$ tags for each node $i$. In Fig.\ref{fig:num_categs}a 
we display the density distribution of $n_i$ in the studied systems, 
whereas Fig.\ref{fig:num_categs}b shows the average number of tags,
 $\left< n\right>$ as a function of the node degree. Since
 the range of the possible $n$ values is rather wide (especially in case of 
 the MathSciNet), we used exponentially increasing bin sizes
 in Fig.\ref{fig:num_categs}a. The decay
of the distributions towards large $n$ values seems exponential.
Concerning the curves shown in Fig.\ref{fig:num_categs}b,
  a plausible hypothesis about 
tagged real networks is that they show tag-assortativity, namely 
links form ties between similar nodes more frequently than at random.
 Therefore, we expect connected nodes to share common tags
 with enhanced probability. Consequently hubs are expected
 to have a larger number of tags than nodes with small degrees, since
 they have to share common attributes with a large number of other nodes. 
Interestingly, in Fig.\ref{fig:num_categs}b the MathSciNet 
 behaves as expected from this point of
 view (with a monotonously increasing $\left< n\right>(k)$ curve), 
 whereas the MIPS network and the Wikipedia do not. For both networks,
$\left< n\right>(k)$ 
 is increasing at small degrees, then in case of the MIPS network it
 saturates, whereas for the Wikipedia it even drops down at high degrees.
This implies that the simple picture shown above, in which the hubs 
correspond to versatile nodes with a large number of different tags does 
not hold in these systems.

\subsection{Tag-induced sub-graphs }
\label{sect:sub_graph_results}

\begin{figure}[t!]
\centerline{\includegraphics[angle=0,width=0.55\columnwidth]{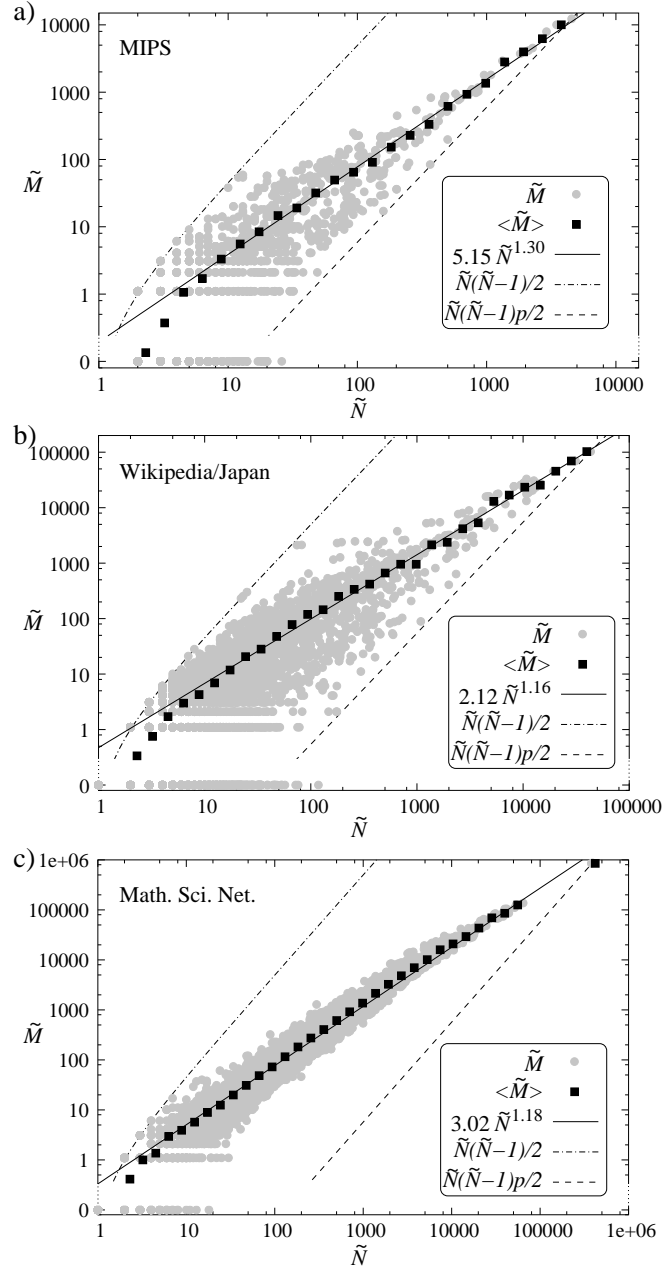}}
\caption{ The scatter-plot 
of the number of links, $\tilde{M}$ versus
 the number of nodes, $\tilde{N}$ in the tag-induced sub-graphs of the 
 different categories (gray symbols) for the MIPS network (a), 
the Wiki-Japan network (b) and
 the MathSciNet. The black symbols show $< \tilde{M}>$, whereas
 the solid lines correspond to the best power-law fit 
to $< \tilde{M}>(\tilde{N})$. 
The dot-dashed lines and the dashed lines in each plot show the upper
 bound in $\tilde{M}$ and the expected number of links for a randomly chosen
 nodes respectively.}
\label{fig:categ_scats}
\end{figure}

Due to the size of the entire Wikipedia, we have been able to analyse
only some of its tag-induced sub-graphs, as described in
Sect.\ref{Sect:construct}. To get a better understanding of the
relationship between tag distribution and network topology, it is very
insightful to go further down this line, and compare some of the basic
properties of the tag-induced sub-graphs for every category (all the
way from the root to the leafs) in all of our three networks. The
scatter plots in Fig.\ref{fig:categ_scats}. with gray symbols
depict the link number ($\tilde{M}$) vs. node number ($\tilde{N}$)
relation for each category. $\tilde{M}$ has a maximum of
$\tilde{M}_{\rm max}(\tilde{N})=\tilde{N}(\tilde{N}-1)/2$,
when the sub-graph forms a clique, i.e., each node is linked to all the
others. This upper bound is shown with a dashed-dotted line.
The estimate of the number of links
$\tilde{M}_{\rm rand}(\tilde{N})=
p\tilde{N}(\tilde{N}-1)/2=p\tilde{M}_{\rm max}(\tilde{N})$
between randomly selected $\tilde{N}$ nodes, is also plotted with a
dashed line, where the linkage probability is defined as
$p=M/[N(N-1)/2]$.
According to the scatter plots, in all the three systems the number of
links $\tilde{M}$ in every tag-induced sub-graph (with some exception
at $\tilde{M}=0$) exceeds the number of links $\tilde{M}_{\rm rand}$
expected for a link distribution that is uncorrelated to the tag
distribution. This strongly indicates that the networks under study are
\emph{tag-assortative}.

\begin{figure}[t!]
\centerline{\includegraphics[angle=0,width=0.75\columnwidth]{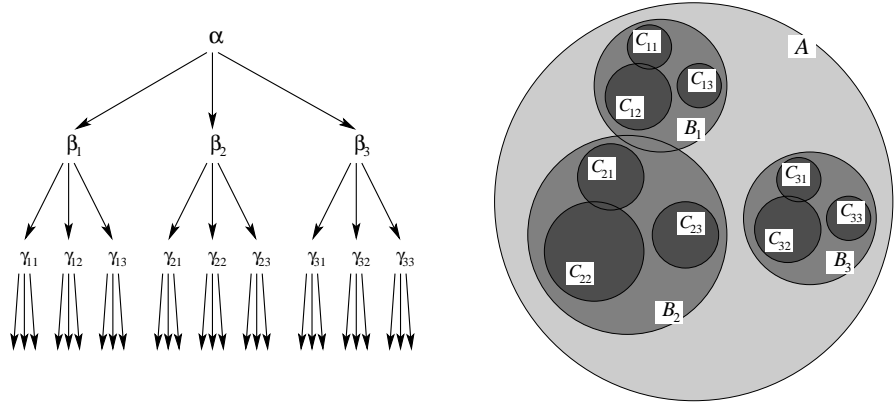}}
\caption{Demonstration of the \emph{self-similar} nature of recursively embedded
tag-induced sub-graphs $C_{ij} \subset B_i \subset A$ generated by
 the DAG of hierarchically organised categories 
$\gamma_{ij} \subset \beta_i \subset \alpha$. The grey level 
is indicative of the link density.
}
\label{fig:Imre_fig}
\end{figure}

An even more intriguing property of the scatter plots is that if the
average number of links $<\tilde{M}>$ are plotted (with black symbols)
as a function of the number of nodes $\tilde{N}$ (using logarithmic
binning), then they strictly follow a power law
$<\tilde{M}>\sim\tilde{N}^\mu$ (solid lines) for several orders of
magnitude (with a deviation only at the smallest sub-graphs). The
\emph{tag-assortativity exponent} $\mu$, defined by this power law,
takes the values of $1.30\pm 0.02$, $1.16\pm 0.02$, and $1.18\pm 0.01$
for the MIPS, the Wiki-Japan, and the MathSciNet networks,
respectively. The physical meaning of this exponent can be demonstrated
by considering the relation between the tag-induced sub-graph of some
category and those of its sub-categories. If the tag-induced (not
necessarily disjoint) sub-graphs of the sub-categories inherit all the
links of the parent category ``homogeneously'' and without having
inter-sub-graph links (i.e., having no links between any pair of
sub-graphs other than those originating in the intersection), then the
number of links corresponding to a sub-category is expected to scale
linearly with the number of its nodes, implying $\mu=1$. If, however,
inter-sub-graph links also appear (cf. Fig.\ref{fig:Imre_fig}), then the number of links
corresponding to a sub-category is expected to drop faster than
linearly, leading to $\mu>1$. Although $\mu<1$ cannot be ruled out (at
least locally, between a particular category and its sub-categories),
it requires very peculiar topologies (e.g. large link density in the
intersection between the tag-induced sub-graphs of two sub-categories)
and, thus, we do not anticipate to obtain such values for real systems.

In brief, a value of $\mu>2$ indicates tag-disassortativity; $\mu=2$
characterises no correlation between tag-similarity and link
distribution (cf. $\tilde{M}_{\rm rand}$); whereas $0<\mu<2$ is the
regime of tag-assortativity with the amendment that $0<\mu<1$ would
represent extreme tag-assortativity. This classification scheme affirms
that the tag-assortativity exponent $\mu$ defined above is indeed an
appropriate quantity for characterising the extent of
tag-assortativity. Our finding that its value for the three
networks we have studied is closer to $1$ than to $2$ suggests that
these networks exhibit a significant tag-assortativity, MIPS
being somewhat less tag-assortative than the other two.

Both the fact that the statistical properties of tag-induced sub-graphs
are similar to those of the entire graph and also the fact that a
single well defined exponent characterises tag-assortativity over
several orders of magnitudes of the sub-graph size imply prominent
\emph{self-similarity} in the structure of tagged networks.
Briefly speaking, the tag-induced sub-graph $A$ of some category
$\alpha$ is related to the tag-induced sub-graphs $B_i \subset A$ of
its sub-categories $\beta_i \subset \alpha$ statistically the same way
as the sub-graphs $B_i$ of categories $\beta_i$ to the tag-induced
sub-graphs $C_{ij} \subset B_i$ of their sub-categories $\gamma_{ij}
\subset \beta_i$, as demonstrated in Fig.\ref{fig:Imre_fig}, 
i.e. both the network
topology and the tag distribution appear to be \emph{scale invariant}.



\subsection{Similarity}
\label{sect:sim_results}
\begin{figure}[t!]
\centerline{\includegraphics[angle=0,width=\columnwidth]{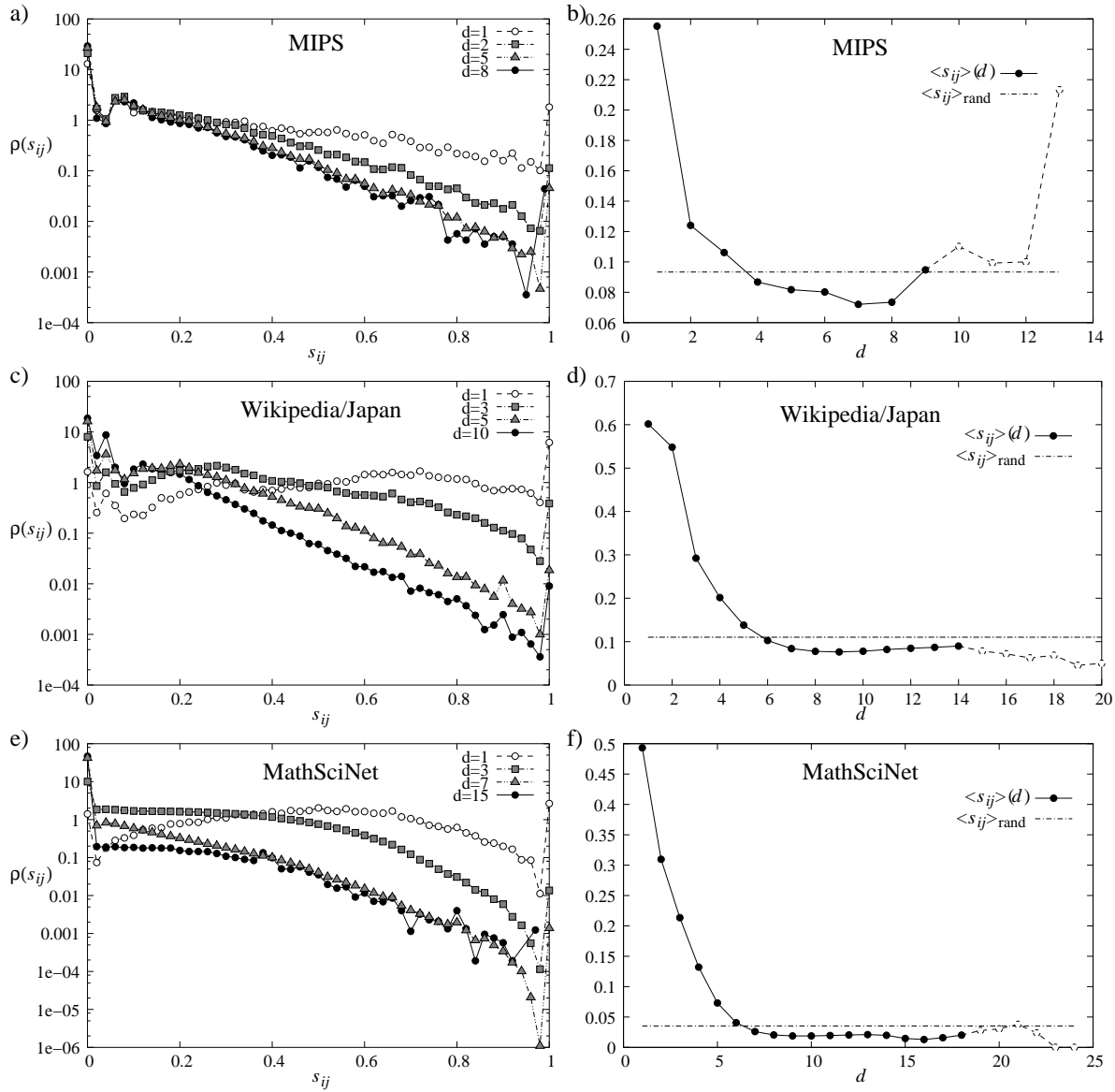}}
\caption{The similarity $s_{ij}$ as a function of the distance between 
the nodes. The density distribution of $s_{ij}$ 
at various distances is plotted on semi-logarithmic scale for 
the MIPS network, 
the Wiki-Japan network
 and the MathSciNet in panels (a), (c) and (e) respectively. The
 corresponding average similarity, $< s_{ij} >$ 
as a function 
of the node distance $d$ is shown in panels (b), (d) and (f). 
The number of 
 pairs at large $d$ becomes small, therefore, the results for 
$< s_{ij} >$ in this regime cannot be trusted. The 
empty symbols and dashed lines indicate that the number of pairs has 
decreased below the total number of links in the network.}
\label{fig:sim_decay_comb}
\end{figure}

The introduction of a similarity measure based on the node tags enable
 us to study other type of relations between the topology and the
 annotations as well. In Fig.\ref{fig:sim_decay_comb} we follow the 
change of the similarity between the nodes with the distance 
in the three networks. The right column of the figure 
shows the density distribution $\rho(s_{ij})$ 
 for $s_{ij}$ obtained from (\ref{eq:max_node_sim}), whereas
 the left column displays the corresponding average similarity,
 $< s_{ij} >$ as a function of the node distance $d$.  
 The $\rho(s_{ij})$ distributions
 are shifted towards lower $s_{ij}$ values with increasing distance
 $d$ between the nodes and accordingly a rapid decreasing tendency
 can be observed in the $<s_{ij}>(d)$ 
function at small distances.
 At medium node distances $< s_{ij}>$ becomes more or less 
 constant, suggesting that the nodes become independent of each other.
In consistency with the results of Sect.\ref{sect:sub_graph_results}., this
 is another indication of \emph{tag-assortativity}: if links were drawn 
between the nodes at random, the $< s_{ij}>$ 
would be independent
 of the distance between the nodes (the $< s_{ij}>(d)$ 
would
 resemble a flat line). The prominent peak at distance $d=1$ signals that
neighbouring nodes are much more similar to each other than at larger
 distances and much more similar to each other than at random as well.

At large node distances the number of pairs is rapidly decreasing 
(i.e., at the possible maximum distance only a few pairs of nodes can
 contribute to $< s_{ij} >$). 
To indicate that the number of 
samples in this regime is not enough for a significant statistical analysis,
we changed the filled symbols (and solid lines) to empty symbols (and
 dashed lines) in Figs.\ref{fig:sim_decay_comb}b,\ref{fig:sim_decay_comb}d and 
\ref{fig:sim_decay_comb}e. Interestingly, for the MIPS network 
$< s_{ij}>(d)$ becomes increasing in this region, reaching a
 value at the maximal distance $d_{\rm max}$ almost as high as at $d=1$. 
 However, this is due  the fact that the five nodes making up the pairs 
 at $d_{\rm max}$ happen to be more similar to a randomly chosen node 
than average.
(Nodes having a couple of non-specific
 tags can be indeed quite similar to the majority of the nodes). 
Since the number
 of pairs at large $d$ is small, the contribution 
 from these nodes is significant, and $< s_{ij} >$ 
becomes larger
 than at medium $d$, where the vast number of other nodes counter balance
 this distortion.

\subsection{Node uniqueness}
\begin{figure}[t!]
\centerline{\includegraphics[angle=0,width=0.6\columnwidth]{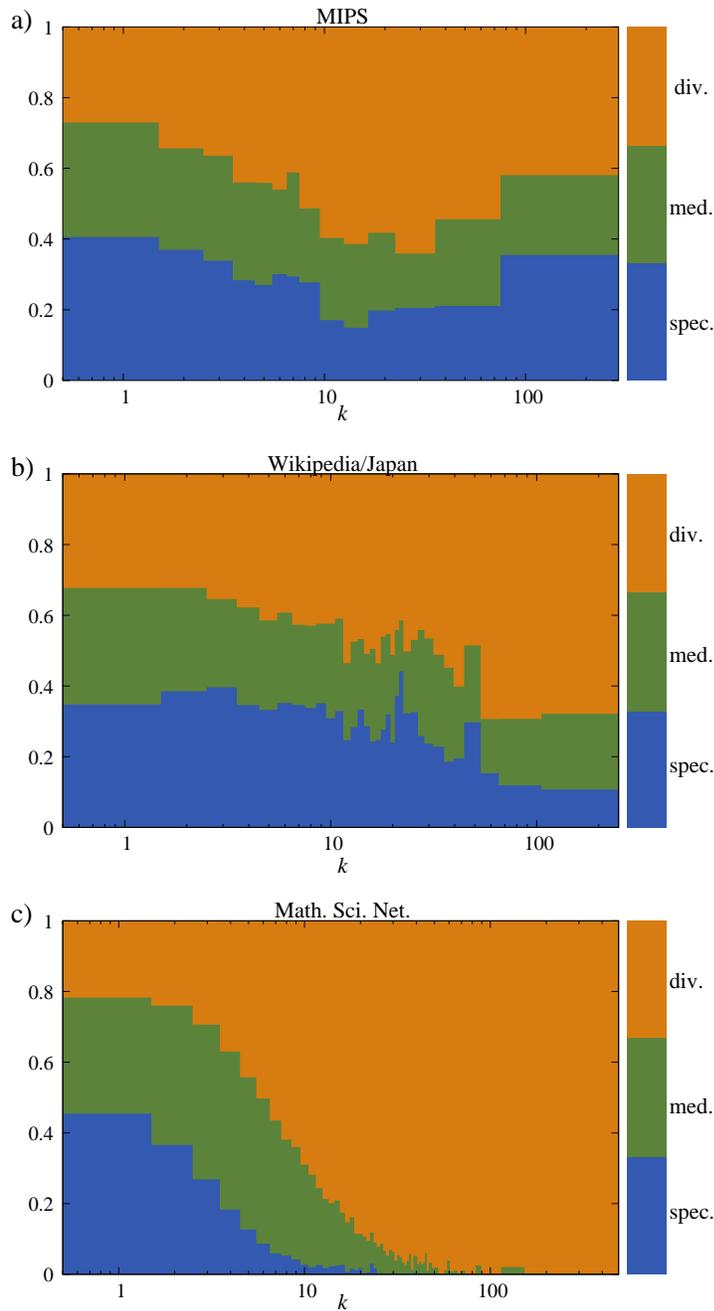}}
\caption{ The participation ratio of the nodes in the three node uniqueness
 classes as a function of the node degree $k$ for the MIPS network (a), the
 Wikipedia/Japan network (b) and the MathSciNet (c).}
\label{fig:K_div_degs}
\end{figure}
We now move on to the investigation of the node uniqueness,
 defined in (\ref{eq:K}). Our main interest concerns the dependence
 of $u$ on the node degree. We divide the nodes into three classes
of equal size depending on their $u$ value: \emph{specific} nodes
 have relatively high $u$ (marked by either a rare label or a few
 closely related rare labels), \emph{medium} nodes have a $u$ value
 around the average, whereas \emph{diverse} nodes have a relatively
 low $u$ value (marked by frequent or un-related labels). In
 Fig.\ref{fig:K_div_degs}. we show the participation ratio of the
 nodes in the three classes as a function of the node degree $k$.
Again, the three systems show different behaviour. In case of the 
 Wikipedia and the MathSciNet, the ratio of diverse nodes is increasing
 monotonously with the node degree. This tendency is very pronounced 
in the latter network (Fig.\ref{fig:K_div_degs}c), where in fact all hubs 
 are classified as
 diverse above a certain degree. This is 
 consistent with the steady increase in the average number of tags as a
 function of the node degree in Fig.\ref{fig:num_categs}b (square symbols):
 for nodes with $n\sim10^2$ tags we expect to find at least a few pairs
 of rather un-related categories resulting in a low $u$ value. 
 In contrast, for the MIPS network,
 the monotonous increase in the ratio of diverse nodes with the node
 degree is followed by a sudden drop at the largest
 degrees. This means that a significant portion of the hubs in this network
 have rather specific functions.

\section{Summary and conclusion}
\label{Sect:summary}

We studied the basic statistical properties of tags in real networks, with
 an interest in the relation between the topology and the tag distribution.
 We found that the investigated systems show universal features in
 some aspects with interesting differences from other perspectives. At small
 and intermediate degrees the average number of tags per node is 
 increasing with the degree, and accordingly the
 node uniqueness 
 is decreasing. For the
 MathSciNet this tendency is prolonged 
 in the high degree regime as well. In contrast, the number of tags on
 the hubs in the MIPS network drops down and simultaneously the ratio of 
 nodes with large uniqueness becomes increasing. The behaviour of the
 English Wikipedia is somewhere in between: the number of tags 
 saturates for the hubs and the further increase in the ratio of nodes with low
 uniqueness is marginal. This comparison reflects the difference in the
 behaviour of hubs in these networks: the hubs of the MathSciNet 
 are very versatile with huge amounts of different tags and 
 low values of uniqueness, 
whereas in the MIPS network a significant portion of the hubs
 correspond to proteins with rather specific functions. 

We introduced the tag-similarity of nodes, which (in contrast to the 
usual structural similarity) can be calculated independently of the 
 graph topology, and is based on the set of tags associated with the nodes. 
According to our results, the studied real networks show tag-assortativity:
the similarity is decreasing with the node distance at
 small range and reaches a minimum at medium distances. In other words,
 tag-similar nodes are linked with each other at higher probability than
 at random. The tag-assortativity is supported by the investigation of 
 the tag-induced sub-graphs as well, since the number 
 of links between the nodes sharing a given tag is always larger than 
(or at least equal to) the number of links expected at random.

An even more interesting property of the tag-induced sub-graphs is that the
average number of their links follow a power law as a function of
 the number of their nodes for several orders of magnitude. 
The magnitude of the 
\emph{tag-assortativity exponent} $\mu$, defined by this power law is
in close relation with the tag-assortativity property of the network:
a value of $\mu>2$ indicates tag-disassortativity; $\mu=2$
characterises no correlation between tag-similarity and link
distribution; whereas $0<\mu<2$ is the
regime of tag-assortativity (with $0<\mu<1$ representing 
extreme tag-assortativity). The tag-assorativity exponent was
slightly above 1 for all studied networks in our case.

The above scaling also reveals that the structure of the 
studied tagged networks is \emph{self-similar}. This is supported by the
 fact that the statistical properties of tag-induced sub-graphs
are similar to those of the entire graph. This means that in
 the statistical sense, the network is related to a sub-graph induced by
 a given category $\alpha$ in the same way as this sub-graph is related 
 to the tag-induced sub-graph of a descendent of $\alpha$, 
i.e. both the network topology and the tag distribution are
\emph{scale invariant}.

\section*{Appendix}
\label{sec:app}

\subsection*{Preparing a Directed Acyclic Graph (DAG) from the
category hierarchy of the English Wikipedia}
\label{subsec:wikcategdag}

In Wikipedia the classification terms of each page (appearing at the
bottom of the page) are called categories and are arranged into a
hierarchy, i.e. a directed network where a more general term is
connected to each of its child terms via a directed link. It is
important to note that this directed graph contains cycles (loops): a
closed path of nodes where each node (a category) is a sub-category of
the previous one and the first is a sub-category of the last. Many of
these loops are short and are made up of a small group of synonymous
terms, e.g., the categories Hindustani and Urdu are very closely
related and are both sub-categories of the other. An example for a
longer loop is Education: Social sciences: Academic disciplines:
Academia: Education, and a loop of length $22$ has been found, too,
in the English Wikipedia \cite{wikcateg}.

Loops in the category hierarchy can confuse both readers and search
engines, and prohibit a tree-based semantic analysis of annotations.
For example, with loops it would impossible to identify the closest
common ancestor(s) of two arbitrary terms and decide their level of
relatedness. To delete all loops from the hierarchy of Wikipedia
categories, first we devised an algorithm eliminating all loops from
a generic directed network by sequentially removing single directed
links and modifying the directed network by the smallest possible
amount. Then, we applied the algorithm to the directed network
defined by the category hierarchy of the English Wikipedia.

The algorithm can be applied to an arbitrary directed network (nodes
connected with directed links) and it has two parts. First, it
identifies the ''loop sub-graph'' of the full directed graph, the set
containing precisely the directed links of all loops. This is
achieved by an iteration where in each step all directed links are
removed that have either a start node that is a source (no incoming
link) or an end node that is a drain (no outgoing links). Neither of
these two node types (source and drain) can be in a loop. Repeating
this removal step until at least one node is removed lead to a
sub-graph containing precisely the loops of the full graph. Note that
the loop sub-graph may have more than one graph component.

The second step of the algorithm identifies a set of directed links
($L$) whose removal from the loop sub-graph eliminates all of its
loops. As the loop sub-graph is by definition the set of loops of the
original graph, removing the same directed links from the full graph
will eliminate its loops. We selected the set of removed links, $L$,
with the goal to modify the full graph by the smallest possible
amount. This concerns not only the size of $L$ (the number of links
removed), but also selecting links with the smallest significance as
viewed from the full graph. Turning back to one of the above
examples, one has to decide which of the two directed links ''Urdu is
a sub-category of Hindustani'' or ''Hindustani is a sub-category of
Urdu'' is less relevant from the point of view of the entire directed
network. More generally, suppose that in a (directed) network the
directed links A$\rightarrow$B and B$\rightarrow$A are both present.
To eliminate the loop $A\rightarrow$B$\rightarrow$A, one of the two
links has to be removed.

To decide which of the two links is less significant, consider
another example. In a directed network with the four links
M$\rightarrow$A, A$\rightarrow$B, B$\rightarrow$A and
B$\rightarrow$N, the link A$\rightarrow$B is more important than
B$\rightarrow$A, because it is contained by a long continuous path,
M$\rightarrow$A$\rightarrow$B$\rightarrow$N. On the other hand,
B$\rightarrow$A points in the opposite direction, thus, it is likely
to be a ''side effect''. The difference between these two links can
be measured. The number of point-to-point shortest directed paths
passing through A$\rightarrow$B is larger ($3$: M$\rightarrow$N,
A$\rightarrow$N and A$\rightarrow$N) than the number of those
containing B$\rightarrow$A (only $1$: B$\rightarrow$A). In a directed
network the number of shortest paths passing through a given
(directed) link is called the directed betweenness centrality of that
link. Multiple shortest paths between two nodes are accounted for by
weighting, see e.g., Ref.~\cite{Dorog_book} for the undirected case.
Based on the above observation, we quantified the significance of
each directed link by its directed betweenness centrality, ${\mathcal
B}$, as measured in the full network.

Now let us return to the second part of the algorithm starting from
the loop sub-graph. Knowing ${\mathcal B}$ of each link in this
sub-net, we can select and remove the least important link, i.e. the
one with the lowest ${\mathcal B}$ value. This link removal may
produce source nodes (only outgoing links) and drain nodes (only
incoming links). Again we iteratively remove links not contained by
loops until the remaining network ''melts down'' to the set of
remaining loops. We repeat this step -- deleting the link with
smallest ${\mathcal B}$ and then iteratively removing all non-loop
links -- until no more links remain. We save the set of removed
links, $L$, and remove the same set of links from the full graph to
eliminate all of its loops by modifying it by the smallest possible
amount.

The full category hierarchy of the English Wikipedia (Oct/17/2007
version) contains $265\, 432$ nodes (categories) and $543\, 722$
directed links (category - sub-category connections). The loop
sub-graph has $4\, 980$ nodes and $13\, 164$ (directed) links. The
total number of removed links was $|L| = 3\, 977$. Data together
and processing programs can be downloaded from the website {\tt 
http://CFinder.org 
}.

\section*{Acknowledgment}
We thank E. Gabrilovich for public domain pre-processing software of
Wikipedia. This work was supported by the Hungarian National Science
Fund (OTKA K68669, K75334 and T049674), the National Research
 and Technological Office (NKTH, CellCom RET, Textrend) and the J\'anos 
Bolyai Research Scholarship of the Hungarian Academy of Sciences.

\section*{References}

\bibliographystyle{unsrt}
\bibliography{annot_art}

\begin{thebibliography}{10}

\bibitem{Laci_revmod}
R.~Albert and A.-L. Barab{\'a}si.
\newblock Statistical mechanics of complex networks.
\newblock {\em Rev.\ Mod.\ Phys.}, 74:47--97, 2002.

\bibitem{Dorog_book}
J.~F.~F. Mendes and S.~N. Dorogovtsev.
\newblock {\em Evolution of Networks: From Biological Nets to the Internet and
  WWW}.
\newblock Oxford University Press, Oxford, 2003.

\bibitem{Watts-Strogatz}
D.~J. Watts and S.~H. Strogatz.
\newblock Collective dynamics of 'small-world' networks.
\newblock {\em Nature}, 393:440--442, 1998.

\bibitem{Faloutsos}
M.~Faloutsos, P.~Faloutsos, and C.~Faloutsos.
\newblock On power-law relationships of the internet topology.
\newblock {\em Comput. Commun. Rev.}, 29:251--262, 1999.

\bibitem{Laci_science}
A.-L. Barab{\'a}si and R.~Albert.
\newblock Emergence of scaling in random networks.
\newblock {\em Science}, 286:509--512, 1999.

\bibitem{Boccaletti}
S.~Boccaletti, V.~Latora, Y.~Moreno, M.~Chavez, and D.-U. Hwang.
\newblock Complex networks: Structure ad dynamics.
\newblock {\em Physics Reports}, 424:175--308, 2006.

\bibitem{Jeong_metabolic}
H.~Jeong, B.~Tombor, R.~Albert, Z.~N. Oltvai, and A.-L. Barab\'asi.
\newblock The large-scale organization of metabolic networks.
\newblock {\em Nature}, 407:651--654, 2000.

\bibitem{Ravasz02}
E.~Ravasz, A.~L. Somera, D.~A. Mongru, Z.~N. Oltvai, , and A.-L. Barab\'asi.
\newblock Hierarchical organization of modularity in metabolic networks.
\newblock {\em Science}, 297:1{55}1--1{55}5, 2002.

\bibitem{Han04}
J.-D.~J. Han, N.~Bertin, T.~Hao, D.~S. Goldberg, G.~F. Berriz, L.~V. Zhang,
  D.~Dupuy, A.~J.~M. Walhout, M.~E. Cusick, F.~P. Roth, and M.~Vidal.
\newblock Evidence for dynamically organized modularity in the yeast
  protein\u2013protein interaction network.
\newblock {\em Nature}, 430:88--93, 2004.

\bibitem{Alon_1}
R.~Milo, S.~Shen-Orr, S.~Itzkovitz, N.~Kashtan, D.~Chklovskii, and U.~Alon.
\newblock Network motifs: Simple building blocks of complex networks.
\newblock {\em Science}, 298:824--827, 2002.

\bibitem{Alon_2}
R.~Milo, S.~Itzkovitz, N.~Kashtan, R.~Levitt, S.~Shen-Orr, I.~Ayzenshtat,
  M.~Sheffer, and U.~Alon.
\newblock Superfamilies of evolved and designed networks.
\newblock {\em Science}, 303:1538--1542, 2004.

\bibitem{Domany_PRL}
M.~Blatt, S.~Wiseman, and E.~Domany.
\newblock Super-paramagnetic clustering of data.
\newblock {\em Phys.\ Rev.\ Lett.}, 76:3251--3254, 1996.

\bibitem{GN-pnas}
M.~Girvan and M.~E.~J. Newman.
\newblock Community structure in social and biological networks.
\newblock {\em Proc.\ Natl.\ Acad.\ Sci.\ USA}, 99:7{82}1--7{82}6, 2002.

\bibitem{Zhou}
H.~Zhou.
\newblock Distance, dissimilarity index, and network community structure.
\newblock {\em Phys.\ Rev.\ E}, 67:061901, 2003.

\bibitem{Newman_fast}
M.~E.~J. Newman.
\newblock Fast algorithm for detecting community structure in networks.
\newblock {\em Phys.\ Rev.\ E}, 69:06{61}33, 2004.

\bibitem{Radicchi}
F.~Radicchi, C.~Castellano, F.~Cecconi, V.~Loreto, and D.~Parisi.
\newblock Defining and identifying communities in networks.
\newblock {\em Proc.\ Natl.\ Acad.\ Sci.\ USA}, 101:2658--2663, 2004.

\bibitem{Huberman_PNAS}
D.~M. Wilkinson and B.~A. Huberman.
\newblock A method for finding communities of related genes.
\newblock {\em Proc.\ Natl.\ Acad.\ Sci.\ USA}, 101:5{24}1--5{24}8, 2004.

\bibitem{Reichard-Bornhold_PRL}
J.~Reichardt and S.~Bornholdt.
\newblock Detecting {F}uzzy {C}ommunity {S}tructures in {C}omplex {N}etworks
  with a {P}otts {M}odel.
\newblock {\em Phys.\ Rev.\ Lett.}, 93:21{87}01, 2004.

\bibitem{Scott_book}
J.~Scott.
\newblock {\em {S}ocial {N}etwork {A}nalysis: {A} {H}andbook}.
\newblock Sage Publications, London, 2nd edition, 2000.

\bibitem{pnas-suppl}
R.~M. Shiffrin and K.~B{\"o}rner.
\newblock Mapping knowledge domains.
\newblock {\em Proc.\ Natl.\ Acad.\ Sci.\ USA}, 101:5183--5185 Suppl.\ 1, 2004.

\bibitem{Everitt-book}
B.~S. Everitt.
\newblock {\em Cluster Analysis}.
\newblock Edward Arnold, London, 3d edition, 1993.

\bibitem{Knudsen-book}
S.~Knudsen.
\newblock {\em A {G}uide to {A}nalysis of {D}{N}{A} {M}icroarray {D}ata}.
\newblock Wiley-Liss, 2nd edition, 2004.

\bibitem{Newman-europhys}
M.~E.~J. Newman.
\newblock Detecting community structure in networks.
\newblock {\em Eur.\ Phys.\ J.\ B}, 38:321--330, 2004.

\bibitem{CPM_nature}
G.~Palla, I.~Der{\'e}nyi, I.~Farkas, and T.~Vicsek.
\newblock Uncovering the overlapping community structure of complex networks in
  nature and society.
\newblock {\em Nature}, 435:814--818, 2005.

\bibitem{Fortunato_coms_chap}
S.~Fortunato and C.~Castellano.
\newblock {\em Community structure in graphs}.
\newblock Springer, Berlin, 2009.

\bibitem{Fortunato_coms_cikk}
A.~Lancichinetti, S.~Fortunato, and J.~Kert\'esz.
\newblock Detecting the overlapping and hierarchical community structure of
  complex networks.
\newblock arXiv:0802.1218, 2008.

\bibitem{Spirin_PNAS}
V.~Spirin and K.~A. Mirny.
\newblock Protein complexes and functional modules in molecular networks.
\newblock {\em Proc.\ Natl.\ Acad.\ Sci.\ USA}, 100:12{12}3--12{12}8, 2003.

\bibitem{Onnela-taxonomy}
J.-P. Onnela, A.~Chakraborti, K.~Kaski, J.~Kert{\'e}sz, and A.~Kanto.
\newblock Dynamics of market correlations: {T}axonomy and portfolio analysis.
\newblock {\em Phys.\ Rev.\ E}, 68:05{61}10, 2003.

\bibitem{Saramaki_stock_Phisica_A}
T.~Heimo, J.~Saram{\"a}ki, J.-P. Onnela, and K.~Kaski.
\newblock Spectral and network methods in the analysis of correlation matrices
  of stock returns.
\newblock {\em Physica A-Statistical Mechanics and its Applications},
  383:147--151, 2007.

\bibitem{Watts-Dodds}
D.~J. Watts, P.~S. Dodds, and M.~E.~J. Newman.
\newblock Identity and search in social networks.
\newblock {\em Science}, 296:1{30}2--1{30}5, 2002.

\bibitem{group_evolv_nature}
G.~Palla, A.-L. Barab\'asi, and T.~Vicsek.
\newblock Quantifying social group evolution.
\newblock {\em Nature}, 446:664--667, 2007.

\bibitem{Szabo_PRE1}
G.~Szab\'o, J.~Vukov, and A.~Szolnoki.
\newblock Phase diagrams for an evolutionary prisoner's dilemma game on
  two-dimensional lattices.
\newblock {\em Phys.\ Rev.\ E}, 72:04{71}07, 2005.

\bibitem{Szabo_PRE2}
J.~Vukov, G.~Szab\'o, and A.~Szolnoki.
\newblock Cooperation in the noisy case: {P}risoner's dilemma game on two types
  of regular random graphs.
\newblock {\em Phys.\ Rev.\ E}, 73:06{71}03, 2006.

\bibitem{Szabo_Phys_Reports}
G.~Szab\'o and G.~F\'ath.
\newblock Evolutionary games on graphs.
\newblock {\em Physics Reports-Review Section of Physics Letters}, 446:97--216,
  2007.

\bibitem{Mason_nets_in_bio}
O.~Mason and M.~Verwoerd.
\newblock Graph theory and networks in {B}iology.
\newblock {\em IET Systems Biology}, 1:89--119, 2007.

\bibitem{Zhu_nets_in_bio}
X.~Zhu, M.~Gerstein, and M.~Snyder.
\newblock Getting connected: analysis and principles of biological networks.
\newblock {\em Genes \& Development}, 21:1{01}0--1{02}4, 2007.

\bibitem{Aittokallio_nets_in_bio}
T.~Aittokallio and B.~Schwikowski.
\newblock Graph-based methods for analysing networks in cell biology.
\newblock {\em Briefings in Bioinformatics}, 7:243--255, 2006.

\bibitem{Finocchiaro_cancer}
G.~Finocchiaro, F.~M. Mancuso, D.~Cittaro, and H.~Muller.
\newblock Graph-based identification of cancer signaling pathways from
  published gene expression signatures using {P}ub{L}i{M}{E}.
\newblock {\em Nucl.\ Ac.\ Res.}, 35:2{34}3--2{35}5, 2007.

\bibitem{Jonsson_Bioinformatics}
P.~F. Jonsson and P.~A. Bates.
\newblock Global topological features of cancer proteins in the human
  interactome.
\newblock {\em Bioinformatics}, 22:2{29}1--2{29}7, 2006.

\bibitem{Jonsson_BMC}
P.~F. Jonsson, T.~Cavanna, D.~Zicha, and P.~A. Bates.
\newblock Cluster analysis of networks generated through homology: automatic
  identification of important protein communities involved in cancer
  metastasis.
\newblock {\em BMC Bioinformatics}, 7:2, 2006.

\bibitem{Zimmermann_coevlov}
M.~G. Zimmermann, V.~M. Egu\'{\i}luz, and M.~S. Miguel.
\newblock Coevolution of dynamical stats and interactions in dynamic networks.
\newblock {\em Phys.\ Rev.\ E}, 69:065102(R), 2004.

\bibitem{Eguiluz_coevolv}
V.~M. Egu\'{\i}luz, M.~G. Zimmermann, and C.~J. Cela-Conde.
\newblock Cooperation and the emergence of role differentiation in the dynamics
  of social networks.
\newblock {\em Am.\ J.\ Sociol.}, 110:977--1008, 2005.

\bibitem{Watts_science}
G.~Kossinets and D.~J. Watts.
\newblock Empirical analysis of an evolving social network.
\newblock {\em Science}, 311:88--90, 2006.

\bibitem{Ehrhardt_coevolv}
G.~C.~M.~A. Ehrhardt and M.~Marsili.
\newblock Phenomenological models of socioeconomic network dynamics.
\newblock {\em Phys.\ Rev.\ E}, 74:036106, 2006.

\bibitem{Newman_coevolv}
P.~Holme and M.~E.~J. Newman.
\newblock Nonequilibrium phase transition in the coevolution of networks and
  opinions.
\newblock {\em Phys.\ Rev.\ E}, 74:056108, 2006.

\bibitem{Gil_coevolv}
S.~Gil and D.~H. Zanette.
\newblock Coevolution of agents and networks: Opinion sperading and community
  disconnection.
\newblock {\em Phys.\ Lett.\ A}, 356:89--94, 2006.

\bibitem{Vazquez_PRE}
F.~Vazquez, J.~C. Gonz\'alez-Avella, V.~M. Egu\'{\i}luz, and M.~S. Miguel.
\newblock Time-scale competition leading to fragmentation and recombination
  transitions in the coevolution of network and states.
\newblock {\em Phys.\ Rev.\ E}, 76:046120, 2007.

\bibitem{Vazquez_cond_mat}
F.~Vazquez, V.~M. Egu\'{\i}luz, and M.~S. Miguel.
\newblock Generic absorbing transition in coevolution dynamics.
\newblock {\em Phys.\ Rev.\ Lett.}, 100:108702, 2008.

\bibitem{Kozma_coevolv}
B.~Kozma and A.~Barrat.
\newblock Consensus formation on adaptive networks.
\newblock {\em Phys.\ Rev.\ E}, 77:016102, 2008.

\bibitem{Benczik_coevolv}
I.~J. Benczik, S.~Z. Benczik, B.~Schmittmann, and R.~K.~P. Zia.
\newblock Lack of consensus in social systems.
\newblock {\em Europhys.\ Lett.}, 82:48006, 2008.

\bibitem{Lambiotte_tags}
R.~Lambiotte and M.~Ausloos.
\newblock Collaborative tagging as a tripartite network.
\newblock {\em Lect. Notes in Computer Sci.}, 3993:1114--1117, 2006.

\bibitem{Jaccard}
P.~Jaccard.
\newblock Nouvelles recherches sur la distribution florale.
\newblock {\em Bull.\ Soc.\ Vandoise des Sci.\ Nat.}, 44:223--270, 1908.

\bibitem{Resnik}
P.~Resnik.
\newblock Semantic similarity in a taxonomy: an information-based measure and
  its application to problems of ambiguity in natural language.
\newblock {\em J.\ Artif.\ Intel.\ Res.}, 11:95--130, 1999.

\bibitem{Lin}
D.~Lin.
\newblock An information-theoric definition of similarity.
\newblock In {\em Proceedings of the 15th International Conference on Machine
  Learning}, pages 296--304, San Francisco CA, 1998.

\bibitem{Guo}
X.~Guo, R.~Liu, C.~D. Shriver, H.~Hu, and M.~N. Liebman.
\newblock Assessing semantic similarity measures for the characterization of
  human regulatory pathways.
\newblock {\em Bioinformatics}, 22:967--973, 2006.

\bibitem{Schlicker}
A.~Schlicker, F.~S. Domingues, J.~Rahnenf{\"u}hrer, and T.~Lengauer.
\newblock A new measure for functional similarity of gene products based on
  gene ontology.
\newblock {\em BMC Bioinformatics}, 7:302--317, 2006.

\bibitem{graph_colouring}
T.~R. Jensen and B.~Toft.
\newblock {\em Graph Coloring Problems}.
\newblock Wiley-Interscience, New York, 1995.

\bibitem{MIPS}
H.~W. Mewes, S.~Dietmann, D.~Frishman, R.~Gregory, G.~Mannhaupt, K.~Mayer,
  M.~Muensterk{\"o}tter, A.~Ruepp, M.~Spannagl, V.~Stuempflen, and T.~Rattei.
\newblock Mips: Analysis and annotation of genome information in 2007.
\newblock {\em Nucl.\ Acids Res.}, 36:D196--D201, 2008.

\bibitem{GO}
The Gene~Ontology Consortium.
\newblock Gene ontology: tool for the unification of biology.
\newblock {\em Nature Genetics}, 25:25--29, 2000.

\bibitem{Mathscinet}
{\tt http://www.ams.org/mathscinet}.

\bibitem{Eng_wiki}
{\tt http://en.wikipedia.org}.

\bibitem{Zlatic_wikipedia}
V.~Zlati\'c, M.~Bo\v zi\v cevi\'c, H.~{\v{S}}tefan{\v{c}}i\'c, and M.~Domazet.
\newblock Wikipedias: Collaborative web-based encyclopedias as complex
  networks.
\newblock {\em Phys.\ Rev.\ E}, 74:016115, 2006.

\bibitem{Capocci_wiki_PRE}
A.~Capocci, V.~D.~P. Servedio, F.~Colaiori, L.~S. Buriol, D.~Donato,
  S.~Leonardi, and G.~Caldarelli.
\newblock Preferential attachment in the growth of social networks: The
  internet encyclopedia wikipedia.
\newblock {\em Phys.\ Rev.\ E}, 74:036116, 2006.

\bibitem{Capocci_wikipedia}
A.~Capocci, F.~Rao, and G.~Caldarelli.
\newblock Taxonomy and clustering in collaborative systems: The case of the
  on-line encyclopedia wikipedia.
\newblock {\em Europhys.\ Lett.}, 81:28006, 2008.

\bibitem{Rios_branching}
P.~De~Los Rios.
\newblock Power law size distribution of supercritical random trees.
\newblock {\em Europhys.\ Lett.}, 56:898--903, 2001.

\bibitem{Caldarelli_branching}
G.~Caldarelli, C.~Caretta Cartozo, P.~De~Los Rios, and V.~D.~P. Servedio.
\newblock Widespread occurrence of the inverse square distribution in social
  sciences and taxonomy.
\newblock {\em Phys. Rev. E}, 69:035101, 2004.

\bibitem{Caldarelli_taxonomy}
C.~Caretta Cartozo, D.~Garlaschelli, C.~Ricotta, M.~Barth\'elemy, and
  G.~Caldarelli.
\newblock Quantifying the taxonomic diversity in real species communities.
\newblock {\em J.\ Phys.\ A: Math.\ Theor.}, 41:224012, 2008.

\bibitem{wikcateg}
{\tt http://en.wikipedia.org/wiki/Wikipedia:Categorization}.

\end{thebibliography}

\end{document}